\documentclass[reprint,onecolumn]{revtex4-2}

\usepackage{graphicx}
\usepackage{dcolumn}
\usepackage{bm}
\usepackage{amsmath}
\usepackage{amsfonts}
\usepackage{braket}
\usepackage{listings}
\usepackage{xcolor}

\definecolor{codebg}{RGB}{246,248,250}
\definecolor{codeframe}{RGB}{200,200,200}
\definecolor{kw}{RGB}{0,92,197}
\definecolor{str}{RGB}{152,0,0}
\definecolor{com}{RGB}{0,128,0}

\lstdefinestyle{python}{
    language=Python,
    backgroundcolor=\color{codebg},
    basicstyle=\ttfamily\small,
    frame=single,
    framerule=0.5pt,
    rulecolor=\color{codeframe},
    keywordstyle=\color{kw}\bfseries,
    commentstyle=\color{com}\itshape,
    stringstyle=\color{str},
    showstringspaces=false,
    numbers=left,
    numberstyle=\tiny,
    numbersep=8pt,
    tabsize=4,
    breaklines=true,
    breakatwhitespace=true,
    captionpos=b,
    keepspaces=true,
    columns=flexible,
    upquote=true,
    lineskip=-1pt,
}

\begin{document}

\title{Implementation of a quantum linear solver for the Vlasov-Amp{\'e}re equation}

\author{Tomer Goldfriend}
 \altaffiliation{Classiq Technologies.
3 Daniel Frisch Street, Tel Aviv-Yafo, 6473104, Israel.}
 \email{tomer@classiq.io}
\author{Or Samimi Golan}
\author{Amir Naveh}%

\date{\today}

\begin{abstract}
We implement a quantum linear solver for the one-dimensional Vlasov–Amp{\'e}re equation, following the model presented in Novikau et. al.
[I. Novikau, I. Y. Dodin, and E. A. Startsev, J. Plasma Phys. 90, 805900401 (2024)]. We design the relevant block encoding operator with Qmod high-level language, and obtain optimized quantum programs using Classiq synthesis tools. Compared to a rigid baseline implementation, our approach yields a clear reduction in quantum resource requirements.
\end{abstract}

\maketitle

\section{Introduction}

The technological and engineering challenge of achieving long-lived, well-confined fusion reactors is accompanied by heavy computational demands.
These include simulating kinetic models of plasma under non-trivial geometries and boundary conditions, capturing its complex and chaotic behavior,
as well as solving large-scale optimization problems with thousands of variables for the design of magnetic confinement devices, such as stellarators.
Quantum computing might offer new approaches that could help with some of the most demanding parts of fusion simulations.

In this work we focus on the application of quantum algorithms for kinetic models of plasma. We follow Ref.~\cite{Novikau_Dodin_Startsev_2024}
that studied the possibility of an efficient quantum algorithm for modeling linear oscillations and waves in a Vlasov model for plasma. 
We provide an explicit implementation for the quantum algorithm, showcasing how Classiq approach~\cite{classiq} can lead to a significant resource reduction.

We consider a kinetic collisionless plasma model, which couples Maxwell’s equations for the electromagnetic field with the motion of charged particles.
The particle motion is described statistically by the distribution function, $f(\vec{x}, \vec{v}, t)$, 
which gives the probability density of finding a particle at position $\vec{x}$ with velocity $\vec{v}$ at time $t$.
The Maxwell-Vlasov equations are written for $f$, and the electric and magnetic fields $\vec{E}$ and $\vec{B}$:
\begin{equation}
    \frac{\partial f}{\partial t} + \vec{v} \cdot \nabla_{\vec{x}} f + \frac{q}{m} \left( \vec{E} + \vec{v} \times \vec{B} \right) \cdot \nabla_{\vec{v}} f = 0,
\end{equation}
\begin{equation}
    \nabla \cdot \vec{E} = \frac{\rho}{\varepsilon_0}, \qquad 
    \mu_0 \varepsilon_0 \frac{\partial \vec{E}}{\partial t} = -\mu_0 \vec{J} +  \nabla \times \vec{B},
\end{equation}
\begin{equation}
    \nabla \cdot \vec{B} = 0, \qquad -\frac{\partial \vec{B}}{\partial t} = \nabla \times \vec{E},
\end{equation}
\begin{equation}
    \rho(\vec{x}, t) = q \int f(\vec{x}, \vec{v}, t) \, d\vec{v}, \qquad
    \vec{J}(\vec{x}, t) = q \int \vec{v} f(\vec{x}, \vec{v}, t) \, d\vec{v},
\end{equation}
where $\rho$ and $\vec{j}$ are the charge density and current, $e$ and $m$ are the particle charge and mass, and $\epsilon_0$ and $\mu_0$ are the vacuum permittivity and  permeability.

In fusion reactors, such as tokamaks or stellarators, there are external magnetic fields driving the plasma, shaping its confinement and influencing its dynamics.
To demonstrate how quantum algorithms can be applied to plasma physics,
we consider a driven problem, though in a simplified one-dimensional model governed by the Vlasov equation coupled with Amp{\'e}re's law.
In certain regimes, this model can be cast into a linear problem, making it suitable for quantum linear system solvers.
 We emphasize that due to the dependence of the quantum solver on the condition number of the linearized system, which grows with the grid size of discretizing the differential equation, we do not expect any quantum advantage for the one-dimensional system.

The rest of this document is organized as follows:
Sections~\ref{sec:model} and~\ref{sec:qls} recap the modeling and problem setup presented in Ref.~\cite{Novikau_Dodin_Startsev_2024}, and in 
Section~\ref{sec:be} we provide a brief description of our quantum model definition.
Section~\ref{sec:res} demonstrates how the Classiq implementation achieves resource savings over a more rigid baseline, and concluding remarks are presented in Sec.~\ref{sec:discussion}.
A complete details on the implementation of the quantum model, including schematic diagrams and Qmod code snippets are provided in Appendix~\ref{app:implementation}, whereas the full code is given in Ref.~\cite{fusion_paper_code}.

\section{A simplified model - the one-dimensional Vlasov–Amp{\'e}re equation}
\label{sec:model}

The simplistic example of Ref.~\cite{Novikau_Dodin_Startsev_2024} is a driven, one dimensional electron plasma 
\begin{equation}
\label{eq:1d_vlasov}
    \frac{\partial f}{\partial t} + v \cdot \nabla_{x} f + \frac{q}{m} E \cdot \nabla_{\vec{v}} f = F_{\rm ext},
\end{equation}
\begin{align}
    \frac{\partial E}{\partial t} &= -\frac{1}{\varepsilon_0 }  \int v f(x, v, t) \, dv +j^{(s)} ,
    \label{eq:amper}
\end{align}
where $F_{\rm ext}$ is some external drive and $j^{(s)}$ is some prescribed current. 
In Eqs.~\eqref{eq:1d_vlasov}-\eqref{eq:amper} we neglect the magnetic field since the system is one-dimensional,
and we omit an explicit equation for the charge density, as it can be inferred from the current $j$ via the continuity equation.

Ref.~\cite{Novikau_Dodin_Startsev_2024} derived a linearized model and obtained the equations:
\begin{align}
\label{eq:dt_g}
\partial_t g + v \, \partial_x g - E \, \partial_v F &= 0, \\
\partial_t E - \int v g \, \mathrm{d}v &= -j^{(S)},
\label{eq:dt_E}
\end{align}
where $g$ is a small perturbation from a background distribution, $F$:
\begin{equation}
    f(x,v,t) = F(x,v) + g(x,v,t).
\end{equation}
In Eqs. \eqref{eq:dt_g}-\eqref{eq:dt_E},  the time, length, and velocity are scaled with the 
\begin{align}
(\text{plasma frequency})^{-1}: \omega^{-1}_p &= \sqrt{\frac{m_e}{4\pi e^2 n_{\rm ref}}} \\
\text{Debye length: } \lambda_{D} &= \frac{v_{\rm th}}{\omega_p}, \\
\text{thermal velocity: } v_{\rm th} &= \sqrt{\frac{T_{\rm ref}}{m_e}},
\end{align}
respectively, where $e$ and $m_e$ are the electron charge and mass, and $n_{\rm ref}$ and $T_{\rm ref}$ are some fixed values of the electron density and temperature.
For the background distribution we take a Maxwellian one, $F(x, v) = \frac{n(x)}{\sqrt{2\pi T(x)}} \exp\left( -\frac{v^2}{2T(x)} \right)$.

\subsection{Boundary value problem}

We focus on the plasma response to a monochromatic external drive at frequency 
$\omega_0$, mirroring the fixed-frequency RF sources commonly employed in fusion devices~\cite{Prater2004}.
\begin{equation}
    j^{(s)}(x,t) = j^{(s)}(x)e^{i\omega_0 t}.
\end{equation}
Thus, we can transform the PDE system of Eqs. \eqref{eq:dt_g}-\eqref{eq:dt_E} into an ODE for two variables, $g(x,v)$ and $E(x)$:
\begin{align}
\label{eq:io0_g}
i\omega_0 g + v \, \partial_x g - E \, \partial_v F &= 0, \\
i\omega_0 E - \int v g \, \mathrm{d}v &= -j^{(S)}.
\label{eq:io0_E}
\end{align}
This coupled system defines the linear problem targeted in our analysis.

\section{Application of Quantum Linear Solver}
\label{sec:qls}

Classical solvers for kinetic equations, especially linearized ones, typically rely on discretizing phase space over a fine grid.
As a result, even modest-resolution simulations require storing and manipulating large matrices, making the computational cost and memory demands significant.
Quantum linear solvers might offer an alternative approach by potentially solving such systems more efficiently in terms of scaling with system size. 
Therefore, we need to discretize our driven kinetic model and reformulate it into a linear system suitable for quantum processing.

First, we rewrite Eqs.~\eqref{eq:io0_g}-\eqref{eq:io0_E} as:
\begin{equation}
    \left(i\omega_0+\mathcal{A}\right)\cdot \begin{pmatrix}
        g \\
        E
    \end{pmatrix} =\begin{pmatrix}
        0 \\
        -j^{(s)}
    \end{pmatrix} , \qquad
    \mathcal{A}\equiv \begin{pmatrix}
        \zeta_{\rm bc} v\partial_x & -\partial_v(F)\cdot \\
        \int  dv\, \cdot v & 0
    \end{pmatrix}
\end{equation}
where we added an artificial boundary condition $\zeta_{\rm bc}$ to avoid incoming waves from the boundaries 
(see Eq.(14) in Ref.~\cite{Novikau_Dodin_Startsev_2024} and Eq.~\eqref{eq:xi} below).
Discretizing the system, i.e., representing phase-space $(x,v)$ as some finite grid and solving for the corresponding fields, leads to a linear equation
that can be solved by matrix inversion:
\begin{equation}
    \left(i\omega_0+A\right)\cdot \vec{\psi} = \vec{b},
\end{equation}
where $A$ is a matrix--- the discretized version of the operators in $\mathcal{A}$, and $\vec{\psi}$ and $\vec{b}$ are
the fields ($g, E$) and the source term ($j^{(S)}$) on the grid points. 

For the discretized system we consider $2^{n_x}\times 2^{n_v}$ grid points for the $(x,v)$ phase space, to represent the domain
$[0,x_{\max}]\times [-v_{\max},v_{\max}]$. Utilizing the bra-ket notation, The statevector $\ket{\psi}$ for the solution is represented by  
\begin{equation}
    \ket{\psi} \equiv \sum\psi_{x,v,e}\ket{x}_{n_x}\ket{v}_{n_v}\ket{E}, \text{  where }
    \left\{\begin{array}{l l}
        \left(\bra{x}_{n_x}\bra{v}_{n_v}\bra{0}\right)\ket{\psi} &= g_{x,v} \\
        \left(\bra{x}_{n_x}\bra{0}_{n_v}\bra{1}\right)\ket{\psi} &= E_{x}
    \end{array}\right. .
    \label{eq:space}
\end{equation}

We define 
\begin{equation}
    A = \begin{pmatrix}
        F & C^{E} \\
        C^{g} & 0
    \end{pmatrix},
    \label{eq:amat}
\end{equation}
where the specific definition of the matrices and how they operate on the effective Hilbert space $\ket{x}_{n_x}\ket{v}_{n_v}\ket{E}$
are described in Appendix~\ref{app:implementation}, in which we provide the explicit quantum implementation.

\subsection{Quantum Singular Value Transform for matrix inversion}

Several quantum routines can solve linear systems, ranging from Harrow, Hassidim, and Lloyd (HHL) algorithm~\cite{HHL} to adiabatic schemes~\cite{adiabatic} and signal-processing approaches~\cite{Martyn2021}, each with its own dependence on the condition number $\kappa$ and target error $\epsilon$. As most of these techniques are based on the block encoding framework (see next paragraphs), we use Quantum Singular Value Transformation~\cite{Martyn2021} (QSVT) as a representative example and focus on building the required block encoding.

QSVT is a framework that embeds a target matrix $B$ 
into a larger unitary and then applies a controlled sequence of single-qubit rotations that implements an arbitrary polynomial $p(\sigma)$
on its singular values. Choosing $p(\sigma)\propto \sigma^{-1}$ over the relevant spectral range turns
this procedure into a quantum matrix-inversion primitive. 
The query and gate complexity of QSVT linear solvers scales polylogarithmically in the dimension and only polynomially in the condition number~\cite{Martyn2021}. We note 
that here the condition number refers to $s/\lambda_{\min}$ with $\lambda_{\min}$ being 
the minimal eigenvalue and $s$ is some prefactor, originating in embedding the problem into a quantum operation, see Eq.\eqref{eq:be_def}.

A QSVT model involves: (1) calculating the rotation angles, (2) preparing an initial state $\ket{\psi_{\rm rhs}}$
on which we would like to apply the inverted matrix, and (3) "Completing" $B$ into a larger unitary matrix, known as {\it block encoding}. This 
document focuses on part (3), which typically dominates the overall resource requirements.

Block encoding of a matrix $B$ refers to the embedding
\begin{equation}
    U_{B,s} \equiv \begin{pmatrix}
        B/s & * \\
        * & *
    \end{pmatrix},
    \label{eq:be_def}
\end{equation} 
where $U_{B,s}$ is unitary and $s$ is some scaling factor. 
The idea is that if we define our Hilbert space as a product of "data" variable and a "block" variable, $\ket{\phi} = \ket{\rm data}\ket{b}$, 
then restricting to the space of $\ket{b}=0$ we have:
\begin{equation}
U_{B,s} \ket{\rm data}\ket{0} = B/s\ket{\rm data}\ket{0} + \text{garbage}.
\end{equation}

\section{Block encoding of $A$}
\label{sec:be}

Our main goal is thus to present an efficient implementation of $U_{i\omega_0+A, s}$, following which, one can apply QSVT 
to invert the the matrix $i\omega_0+A$ and solve the original problem.
We construct the block encoding using standard quantum algorithmic techniques, including the linear combination of unitaries (LCU) method,
and block-encoding arithmetic operations such as inner and outer products. 
Detailed circuit constructions and logic are provided in Appendix~\ref{app:implementation}. 
For all quantum functions, we rely on efficient quantum implementations, except for two primitives: preparing a state 
$\ket{0}\rightarrow \sum_v\eta(v)\ket{v}$, and amplitude assignment $\ket{v}\ket{0}\rightarrow \sum_v\eta(v)\ket{v}\ket{1}+\sqrt{1-\eta^2(v)}\ket{v}\ket{0}$,
with linear, Gaussian, or linear times Gaussian $\eta(v)$. For those, we adopt simple naive implementations for clarity and ease of demonstration.
More efficient methods exist in the literature~\cite{Gonzalez_et_al2024,gaussian2014,Gilyen_etal2019}, and can be included in a future work. (In addition, for small problem sizes, non-scalable methods often yield lower gate depth and qubit counts; the crossover point--- where scalable methods become more efficient--- typically occurs at higher qubit numbers.). 

We note that the original work by Novikau et al.~\cite{Novikau_Dodin_Startsev_2024} also presented the block-encoding construction, albeit using low-level details and techniques. In contrast, the Qmod approach provides a more transparent formulation, highlighting the connection between the classical matrix and its quantum data representation.

The construction of the block encoding using the high-level Qmod language  involves $n_x+n_v+1$ logical qubits for the data variable and additional 8 logical qubits for the block variable.
In the next section we preset how synthesizing high-level description results in significant resource reductions.

\section{Resource reduction using Classiq compiler}
\label{sec:res}

We compare our synthesized quantum program to the one obtained by a fixed, non-flexible Qiskit implementation.
We consider a single QSVT step, applied on $U_{i\omega_0+A,s}$. This model contains one call for $U_{i\omega_0+A,s}$ and its inverse, together with
two calls for rotating an extra qubit, controlled over the block variable~\cite{Martyn2021}. The model thus operates on $n_x+n_v+10$ logical qubits.
As can be seen from the code provided in Appendix~\ref{app:implementation}, the Qmod description contains, explicitly, the following quantum primitives and functions: 
control and inverse operations, inplace addition and xor, simple logical arithmetic operations,
arbitrary state preparation and amplitude assignment, single qubit gate operation, and the CX two qubit gate.
Qiskit has a similar functionality, except for amplitude assignment, logical arithmetic, and in-place addition by a constant.
We implemented those operations in Qiskit, to follow the low-level implementation in Classiq.

\begin{figure}[h!]
\centering
\includegraphics[width=0.8\linewidth]{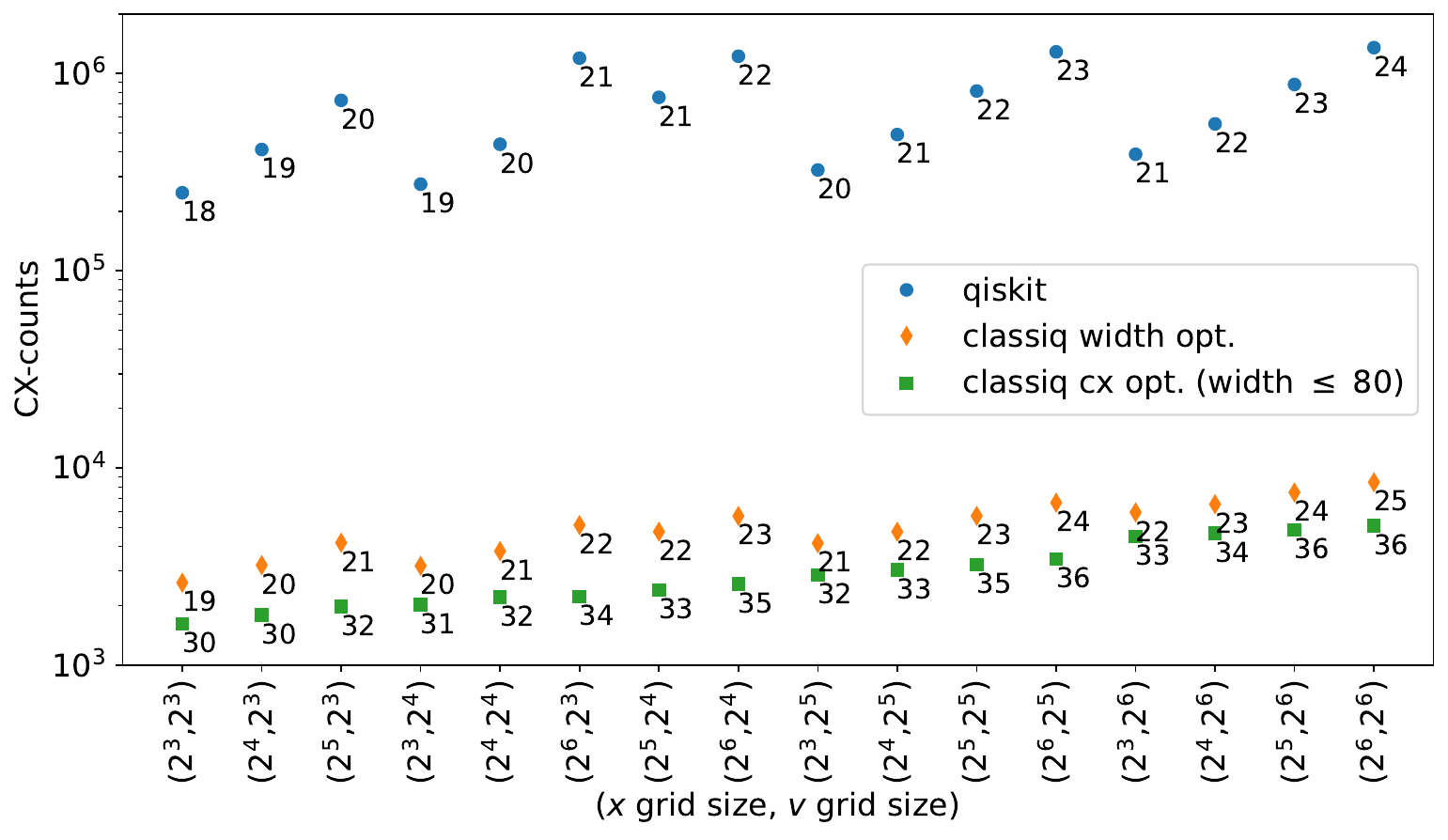}
\caption{The total number of CX gates in a single QSVT step, applied on the block encoding unitary $U_{i\omega_0+A,s}$,
for different problem sizes. The number on each point represents the total width of the quantum program.}
\label{fig:comparison_graph}
\end{figure}

While Qiskit construction is fixed, the Qmod modeling allows smart control operations, as well as automated determination between different multi-control
and adder implementations~\footnote{Classiq built-in adder supports two different implementation, QFT based adder and ripple-carry adder. For 
Qiskit, we fixed the QFT based one.}. We choose to synthesize the Qmod model for two different scenarios: (1) optimization over quantum program width, and
(2) optimization over CX-counts with constraining over the maximal width. Figure~\ref{fig:comparison_graph} shows the CX-counts as a function of the problem size, $(n_x, n_v)$.
The number on each point represents the width of the quantum program. Clearly, Classiq approach can reduce the CX-counts by two orders of magnitudes, compared to
the baseline Qiskit result.

For more complex and realistic problems, e.g., in higher dimensions, for which the quantum solver might provide a speedup over a classical one, the advantage of Classiq approach should be even more pronounced. 

\section{Discussion}
\label{sec:discussion}

This work explores the implementation of a quantum linear solver for the linearized Vlasov-Amp{\'e}re equation, focusing on block-encoding the corresponding classical matrix via a quantum function. We have addressed the one-dimensional case, providing an explicit Qmod implementation that demonstrates how high-level synthesis can yield significant resource reductions.

The specific problem considered here is not expected to exhibit quantum advantage, primarily due to the scaling of the matrix condition number with problem size. However, it may offer polynomial speedups in higher dimensions, as is typical for Finite Element Method problems~\cite{Montanaro&Pallister2016}. We emphasize that the implementation presented does not introduce any additional overhead to the condition number from quantum data embedding; specifically, the factor $s$ does not grow with the problem size (see Eq.~\eqref{eq:final_lcu}).

We hope that the well-structured implementation and demonstrated resource efficiency in this work will be valuable for future, more complex applications of quantum solvers that rely on encoding classical matrices for quantum algorithms.

\begin{acknowledgments}
We thank Ilya Dodin and Ivan Novikau for helpful discussions.
\end{acknowledgments}

\appendix
\section{Implementation with Classiq}
\label{app:implementation}
Below we provide details on the block encoding implementation: presenting the matrix operations explicitly, 
explaining how to construct the corresponding quantum functions, and providing model diagrams and Qmod code snippets. 
We consider the operation of the matrix $A$ in Eq.~\eqref{eq:amat}  on the effective Hilbert space
defined in Eq.~\eqref{eq:space}.
The full code, including the application of a matrix inversion with QSVT, is given in Ref.~\cite{fusion_paper_code}.

\subsection{The upper left matrix $F$}

The upper left part of $A$ refers to the discrete operation
$F=\zeta_{\rm bc}\cdot (\vec{v}\otimes \mathbb{I}) \cdot(\mathbb{I}\otimes\nabla_x)$, given by the product of the following matrices:
\begin{equation}
  \zeta_{\rm bc}\cdot \left(|x\rangle |v\rangle \right) = \left( \left\{ \begin{array}{ll}
       0, & \quad x =0,\, v>0\\
       0, & \quad  x = 2^{n_x}-1,\, v<0\\
       1, & \quad \text{otherwise}
  \end{array} \right. \right)|x\rangle |v\rangle, 
  \label{eq:xi}
\end{equation}
\begin{equation}
  (\vec{v}\otimes \mathbb{I} )\cdot \left(|x\rangle |v\rangle \right) =  v|x\rangle |v\rangle,
\end{equation}
\begin{equation}
  \nabla_x \left(|x\rangle |v\rangle \right) = \frac{1}{2\Delta x}\left( \left\{ \begin{array}{ll}
       -3|0\rangle + 4|1\rangle -|2\rangle , & \quad x = 0 \\
       3|2^{n_x}-3\rangle  -4|2^{n_x}-2\rangle +|2^{n_x}-1\rangle, & \quad x =  2^{n_x}-1\\
       |x+1\rangle - |x-1\rangle , & \quad \text{otherwise}
  \end{array} \right.  \right) |v\rangle, 
\end{equation}
where we omit the $\ket{E}$ qubit as this block acts on $\ket{E}=\ket{0}$. 

The first two operations can be block-encoded via  simple, well-known quantum primitives:
$\zeta_{\rm bc}$ is  given by xor-ing the relevant arithmetic condition with a single block variable,
and $(\vec{v}\otimes \mathbb{I})$  refers to a diagonal matrix with the typical "amplitude assignment" operation
$|v\rangle|0\rangle \rightarrow \eta(v)|v\rangle|1\rangle + \sqrt{1-\eta^2(v)}|v\rangle|0\rangle $, with $\eta(v)=v/v_{\max}$ in our case.
For both of these operations we use Classiq built-in arithmetic and numeric assignment~\cite{qmod2025}; see listings~\ref{lst:be_xi}, and~\ref{lst:be_v} (As mentioned in the main text, the amplitude assignment is given by a non-scalable implementation,
yet it is very practical for demonstration. One can devise a scalable block encoding for a diagonal matrix, e.g., following the idea in Ref.~\cite{Gilyen_etal2019}).
We designate the block encoding of those two operations as $U_{\zeta,1}$ and $U_{v,v_{\max}}$, respectively.

\begin{lstlisting}[caption={Block encoding $\zeta_{\rm bc}$.}, label={lst:be_xi}]
@qfunc
def zeta_be(x: QNum, v: QNum, flag: QBit):
    flag ^= ((x==0) & (v>0))
    flag ^= ((x==(2**x.size-1)) & (v<=0)) # instead of doing 'or', take advantage of the mutual exclusiveness of the conditions
\end{lstlisting}
\begin{lstlisting}[caption={Block encoding $ (\vec{v}\otimes \mathbb{I})$.}, label={lst:be_v}]
@qfunc
def v_be(v: QNum, flag: QBit):
    assign_amplitude(v/(2**(v.size-1)),flag)
    X(ind) # apply an X gate since the assignment is for flag=1, whereas block encoding is defined with flag=0.
\end{lstlisting}

\begin{figure}[h!]
\centering
\includegraphics[width=0.8\linewidth]{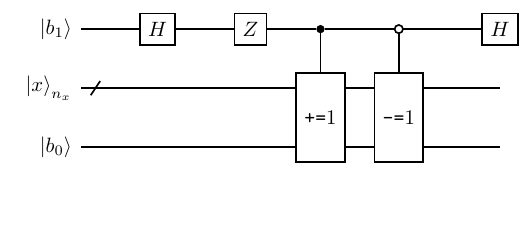}
\caption{Implementation of $U_{{\nabla}_x^{\rm bulk}, \Delta x^{-1}}$, following Ref.~\cite{BE_Sunderhauf}, that provides a general technique for block encoding matrices with fixed diagonals.}
\label{fig:be_dx_bulk}
\end{figure}

Next, we focus on the block encoding for $\nabla_x$, which comprises of two parts, one for the bulk and one for the boundaries:
\begin{equation}
\begin{split}
\nabla_{x} &=  \nabla^{\rm bulk}_{x} + \nabla_{x}^{\rm bc} \\
           &= \nabla^{\rm bulk}_{x} + \nabla_{x}^{\rm bc-up} + \nabla_{x}^{\rm bc-down} \\
           &= \frac{1}{2\Delta x}\left[
\begin{pmatrix}
  0  & 1      &        &        &        \\
 -1  & 0      & 1      &        &        \\
     & -1     & 0      & \ddots &        \\
     &        & \ddots & \ddots & 1      \\
     &        &        & -1     & 0
\end{pmatrix}
+
\begin{pmatrix}
-3  & 4  & -1 & 0 & \cdots & 0 \\
0   & 0  & 0  & \cdots & \cdots & 0 \\
\vdots &    &    & \vdots \\
\vdots &    &    & \vdots \\
0   & 0  & 0  & 0 & \cdots & 0 \\
0   & 0  & 0  & 0 & 0 & 0
\end{pmatrix}
+
\begin{pmatrix}
0  & 0  & 0  & 0 & \cdots & 0 \\
0  & 0  & 0  & \cdots & \cdots & 0 \\
\vdots &    &    & \vdots \\
\vdots &    &    & \vdots \\
0  & 0  & 0  & 0 & \cdots & 0 \\
0  & \cdots & 0 & 3 & -4 & 1
\end{pmatrix}
\right]
\end{split}
\end{equation}
The bulk term can be implemented with two block qubits, following Ref.~\cite{BE_Sunderhauf} (see Eq. (56) therein), where the resulting scaling factor is $\Delta x^{-1}$.
The code for implementing $U_{{\nabla}_x^{\rm bulk}, \Delta x^{-1}}$ and the its schematic description,
are given in listing~\ref{lst:be_dx_bulk} and Fig.~\ref{fig:be_dx_bulk}, respectively.
\begin{lstlisting}[caption={Block encoding ${\nabla}_x^{\rm bulk}$.}, label={lst:be_dx_bulk}]
# derivative along x without boundary conditions
@qfunc
def derivative_dirichlet_be(x: QNum, b1: QBit, b2: QBit):
    extended_qnum = QNum()
    within_apply(
        lambda: bind([x, b1], extended_qnum),
        lambda: lcu(
            [1, -1],
            [
                lambda: inplace_add(-1, extended_qnum),
                lambda: inplace_add(1, extended_qnum),
            ],
            b2,
        ),
    )
\end{lstlisting}

\begin{figure}[h!]
\centering
\includegraphics[width=\linewidth]{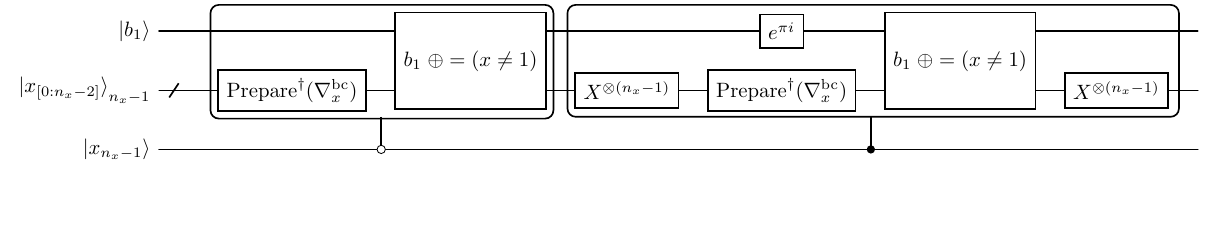}
\caption{Implementation of $U_{{\nabla}_x^{\rm bc}, \alpha\Delta x^{-1}}$. Assuming that
$n_x>2$ we can use $x_{n_x-1}$ for generating the combination of the left and right boundary
conditions, without introducing an extra block qubit and increasing the scaling factor.}
\label{fig:be_dx_bc}
\end{figure}

Next, consider the derivative on the boundaries, which is written as a sum of two matrices, each of which, has a single non-vanishing row with only three
non-vanishing entries. Under the assumption that $n_x>2$, we can implement the two parts on $n_x-1$ qubits, and then control over the last qubit $|x_{n_x-1}\rangle$ as a control variable,
to block encode the sum. Thus, let us focus on the implementation of $U_{\nabla_x^{\rm bc-up}, \alpha\Delta x^{-1}}$,
with some pre-factor $\alpha$ to be determined. Notice that  $\nabla_x^{\rm bc-up}$ is
a product of some matrix whose first row is $(-3, 4, -1,0,\dots, 0)$ and the matrix which has 1 on its first entry and zero elsewhere. 
The former can already be described, up to normalization factor, by the inverse of a state preparation unitary,
and the latter can be block-encoded with a single block qubit $|b\rangle$ using the arithmetic operation $b =b\oplus (x\neq 1)$. 
The fact that block encoding of matrix product is the product of the block encoding of the matrices gives the implementation of $U_{\nabla_x^{\rm bc-up}, \alpha}$,
where the scaling factor $\alpha$ comes from the normalization of the state preparation, $\alpha=|(-3,4,-1)|=\sqrt{26}$.
Finally, the block encoding of $\nabla_x^{\rm bc-down}$ is achieved by adding a minus sign to $U_{\nabla_x^{\rm bc-up}, \alpha\Delta x^{-1}}$,
and transposing the matrix along the opposite diagonal, using a series of $X$ gates. The code that implements this operation and the corresponding
schematic model are given in listing~\ref{lst:be_dx_bc} and Fig.~\ref{fig:be_dx_bc} , respectively. 
\begin{lstlisting}[caption={Block encoding ${\nabla}_x^{\rm bc}$.}, label={lst:be_dx_bc}]
BC_VALUES = np.array([-3/2, 4/2, -1/2,0]) + np.array([0, -1/2, 0,0]) # boundary
BC_VALUES = BC_VALUES/ np.linalg.norm(BC_VALUES) # sqrt(26)
@qfunc
def prepare_bounday_x(x: QArray):
    inplace_prepare_amplitudes(BC_VALUES, 0.0, x[0:2])

@qfunc
def derivative_boundary_left_be(x: QNum, b: QBit):
    invert(lambda: prepare_bounday_x(x))
    b ^= (x != 0)

@qfunc
def derivative_boundary_right_be(x: QNum, b: QBit):
    within_apply(lambda: apply_to_all(X, x), # flip row order
                    lambda: [RY(2*np.pi, b), # phase of -1
                            derivative_boundary_min_be(x, b)])

@qfunc
def derivative_boundaries_be(x: QArray, b: QBit):
    control(x[x.len-1]==0, 
            lambda: derivative_boundary_min_be(x[0:x.len-1], b),
            lambda: derivative_boundary_max_be(x[0:x.len-1], b))
\end{lstlisting}

\begin{figure}
    \centering
    \includegraphics[width=\linewidth]{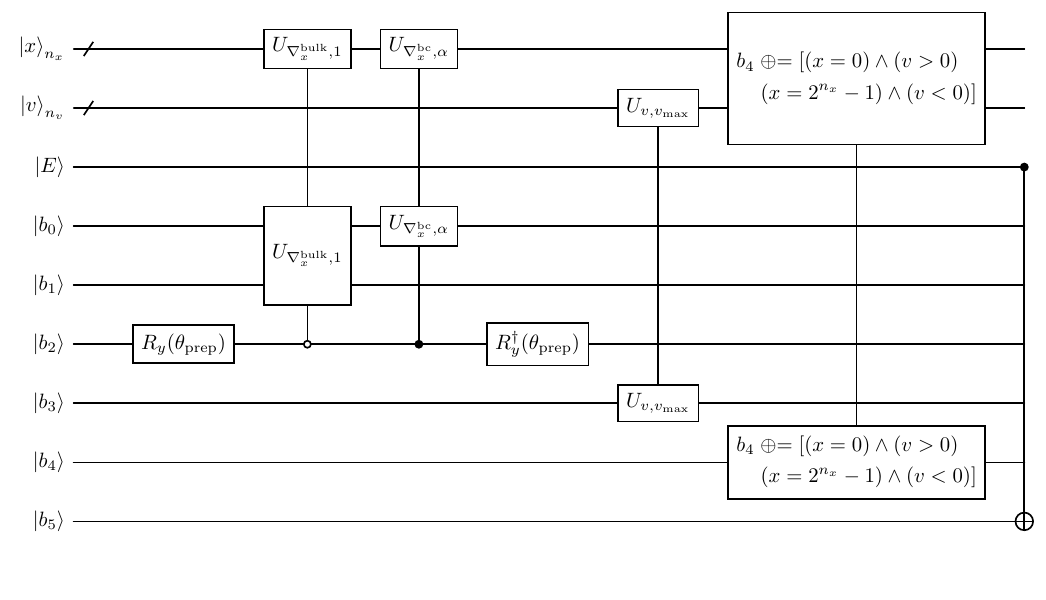}
    \caption{Schematic model for $U_{F,s_F}$, which block encodes the upper left sub-matrix of $A$, with a scaling factor of $v_{\rm max}(1+\alpha)\Delta x^{-1}$.
    The first four operations encode the derivative operator, using prepare and select with $\theta_{\rm prep}=2\arccos\left(\sqrt{\alpha/(1+\alpha)}\right)$,
    and $\alpha=\sqrt{26}$, where the schemes for $U_{\nabla^{\rm bulk}_x,\Delta x^{-1}}$ and $U_{\nabla^{\rm bc}_x,\alpha\Delta x^{-1}}$ are given in Figs.~\ref{fig:be_dx_bulk} and~\ref{fig:be_dx_bc}, respectivaly. The following two operations implement $\zeta_{\rm bc}\cdot (\vec{v}\otimes \mathbb{I})$, and the final CX gate eliminates the lower block on the diagonal, placing $F$ on the upper left side of the larger matrix $A$.}
    \label{fig:be_dx}
\end{figure}

To block encode the complete derivative operator, including the bulk and the boundaries, we apply an LCU:
\begin{equation}
    U_{\nabla_x, (1+\alpha)\Delta x^{-1}} = \frac{\alpha}{1+\alpha}U_{\nabla_x^{\rm bc}, \alpha\Delta x^{-1}}+\frac{1}{1+\alpha}U_{\nabla_x^{\rm bulk}, \Delta x^{-1}}
\end{equation}
This is implemented using the common "prepare and select" technique, involving an extra block qubit. In Fig.~\ref{fig:be_dx} we introduce the schematic model of the full advective
term $F=\zeta_{\rm bc}\cdot (\vec{v}\otimes \mathbb{I}) \cdot(\mathbb{I}\otimes\nabla_x)$. The implementation corresponds to six block qubits, and a scaling factor of $s_F=v_{\rm max}(1+\sqrt{26})\Delta x^{-1}$. 

\subsection{The off-diagonal sub-matrices $C^E$ and $C^g$}

The matrices $C^E$ and $C^g$ apply a simple vector multiplication:
\begin{align}
  C^{E}\cdot \left(|x\rangle |v=0\rangle|E=1\rangle \right) &= -\left(\partial_vF\right) \ket{x}\ket{v}\ket{E=0}, \\
  C^{E}\cdot \left(|x\rangle |v\neq 0\rangle|E\rangle \right) &= 0, \\
  C^{E}\cdot \left(|x\rangle |v\rangle|E=0\rangle \right) &= 0,
\end{align}
\begin{align}
  C^{g}\cdot \left(|x\rangle |v\rangle|E=0\rangle \right) &= -\left(vdv \right) \ket{x}\ket{0}\ket{E=1}, \\
  C^{g}\cdot \left(|x\rangle |v\rangle|E=1\rangle \right) &= 0,
\end{align}
For a given $\ket{x}$, these correspond to matrices with a single column, $-\left(\partial_vF\right)$ for $C^{E}$, and a single
row, $vdv$ for $C^{g}$~\footnote{Notice that the values $\ket{E=1}\ket{v\neq 0}$ are redundant. When running a QSVT solver, we can eliminate them through the projector of the block encoding.}. Therefore, we can follow the same idea for defining $U_{\nabla_x^{\rm bc-up}, \alpha\Delta x^{-1}}$, where now the scaling factors originate in the normalization of the state preparation $\left|\left(\partial_vF\right)\right|$ and $\left|vdv\right|$. 
For completeness, we provide the Qmod implementation in listing~\ref{lst:be_ce_cg}.
We note that here, as opposed to $\nabla_x^{\rm bc-up}$, we need to prepare more complex states. For simplicity we use Classiq built-in arbitrary state preparation function, while an efficient and scalable implementation can be included as well (by calling the function \texttt{prepare\_linear\_amplitudes} from Classiq open library and implementing Gaussian states according to Ref.~\cite{gaussian2014}).

\begin{lstlisting}[caption={Block encoding $C^E$ and $C^g$. V\_MAX and N\_V are global parameters of the problem, standing for $v_{\max}$ and $n_v$ in the text.}, label={lst:be_ce_cg}]
v_amplitudes = np.linspace(-1, 1 - 2 ** (-N_V + 1), 2**N_V) * V_MAX
v_amplitudes = np.roll(v_amplitudes, len(v_amplitudes) // 2) # adjust to signed number
v_amplitudes /= np.linalg.norm(v_amplitudes)
v_H_amplitudes = (
    v_amplitudes
    * np.exp(-(v_amplitudes**2) / (2 * Temperature))
    / (np.sqrt(2 * np.pi * Temperature))
)
v_H_amplitudes /= linalg.norm(v_H_amplitudes) 
c_E_factor = np.linalg.norm(v_H_amplitudes)
c_g_factor = np.linalg.norm(v_amplitudes) * DV

@qfunc
def force_term_be(v: QNum, b: QBit):
    b ^= (v != 0)
    inplace_prepare_amplitudes(v_H_amplitudes, 0, v)

@qfunc
def current_term_be(v: QNum, b: QBit):
    invert(lambda: inplace_prepare_amplitudes(v_amplitudes, 0, v))
    b ^= (v != 0)
\end{lstlisting}

\begin{figure}
    \centering
    \includegraphics[width=\linewidth]{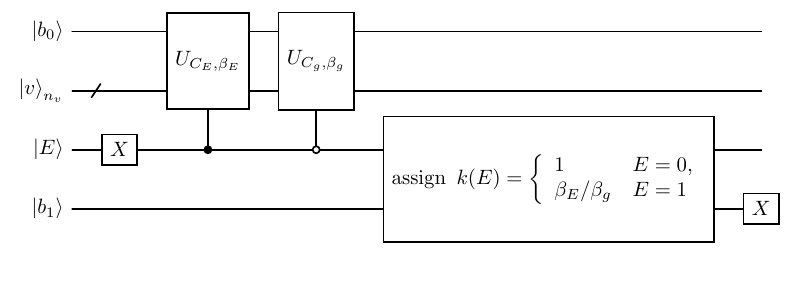}
    \caption{Schematic model for block encoding the off-diagonal matrices of $A$. Each sub matrix is encoded in similar to the boundary terms of $\nabla_x$. This is done with a single block variable and scaling factors $\beta_E=\left|\left(\partial_vF\right)\right|$ and $\beta_g=\left|vdv\right|$, where it is understood that these vectors are discretized according to the $v$ variable with $n_v$ points. The amplitude assignment refers to $\ket{E}\ket{b_1=0}\rightarrow 
    k(E)\ket{E}\ket{1}+\sqrt{1-k^2(E)}\ket{E}\ket{0}$. It is assumed that $\beta_E<\beta_g$, otherwise, we need to assign $\beta_g<\beta_E$ for $E=0$.}
    \label{fig:be_cgce}
\end{figure}

As a final step, we need to insert $C^{E}$ and $C^{g}$ into the upper right and lower left parts of the matrix $A$, 
taking into account their relative prefactor. This can be done by controlling their operation over the $\ket{E}$ variable, together with adding their relevant weight via amplitude assignment on an additional block qubit.
The code for implementing this is given in listing~\ref{lst:be_off_diag}, and the corresponding
schematic diagram is drawn in Fig.~\ref{fig:be_cgce}. This results in a block encoding unitary $U_{C,s_C}$, for 
$C\equiv \begin{pmatrix}
    0 & C^{E}\\
    C^{g}& 0
\end{pmatrix}$, with a block variable of size 2 and a scaling factor of $s_C=\max\left\{\left|vdv\right|, \left|\left(\partial_vF\right)\right|\right\}$.
\begin{lstlisting}[caption={Block encoding both off-diagonal blocks.}, label={lst:be_off_diag}]
@qfunc
def equalize_amplitude(E_field: QNum, ind: QBit, ratio: float):
    """
    Multiply amplitude of |E=1> by ratio, do nothing for |E=0> if ratio <=1
    Multiply amplitude of |E=0> by 1/ratio, do nothing for |E=1> if ratio >1
    """
    amplitudes = np.array([ratio, 1])
    amplitudes /= max(amplitudes)

    assign_amplitudes(subscript(amplitudes, E_field), ind)
    ind ^= 1  # the loaded function is on the |ind=1> state, change to |ind=0>

    
@qfunc
def off_diag_be(E: QBit, v: QNum, b: QArray):
    X(E)
    control(E == 0,
            lambda: force_term_be(v, b[0]),
            lambda: current_term_be(v, b[0]))
    
    # re-weight the diagonals - decrease the term with smaller factor
    equalize_amplitude(E, b[1], c_E_factor / c_g_factor)
\end{lstlisting}

\subsection{Block encoding the Full matrix}
\label{sec:full_be}

We have implemented $U_{F,s_F}$, and $U_{C,s_C}$, with $s_F=v_{\rm max}(1+\sqrt{26})\Delta x^{-1}$, and $s_C=\max\left\{\left|vdv\right|, \left|\left(\partial_vF\right)\right|\right\}$. To construct the block encoding of the full matrix to be inverted, $i\omega_0+A$, we apply
an LCU, with a block variable of size 2:
\begin{equation}
    U_{i\omega_0+A, s} = \frac{s_F}{s}U_{F,s_F}+\frac{s_C}{s}U_{C,s_C}+\frac{\omega_0}{s}iI, \qquad s\equiv s_F+s_C+\omega_0.
    \label{eq:final_lcu}
\end{equation}
The final model has eight block qubits, six from $U_{F,s_F}$ and the additional two from the final LCU
(the block variables of $U_{C,s_C}$ can be shared with the ones of $U_{F,s_F}$, since their operation is mutually exclusive under the "Select" routine). The final scaling factor does not {\it increase} with the system size $2^{n_x}\times 2^{n_v}$, and therefore does not create a infinitesimally vanishing effective condition number.

\begin{figure}[h!]
\centering
\includegraphics[width=0.8\linewidth]{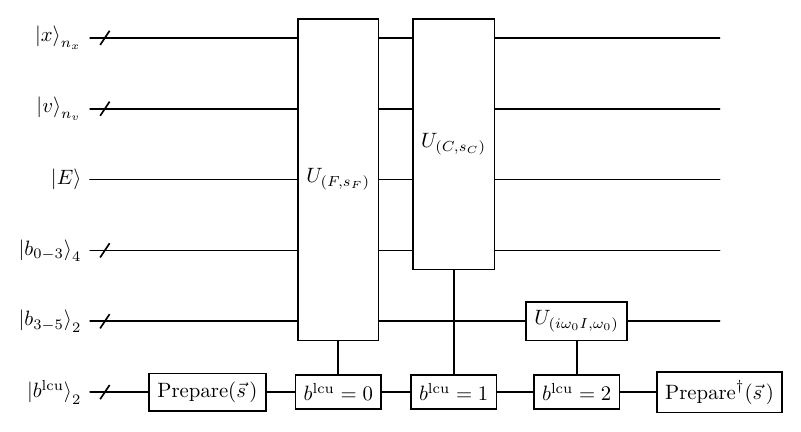}
\caption{Implementation of $U_{i\omega_0+A,s}$, combining all three parts of
the matrix using the LCU technique. The prepared state is $\vec{s}=(s_F, s_C, \omega_0,0)$, containing the scaling factor of each sub-matrix.}
\label{fig:be_full}
\end{figure}

Fig.~\ref{fig:be_full} shows the final layout of the quantum model that implements the block encoding of $i\omega_0+A$. Now, this quantum function can be inserted to a QSVT routine for solving the linear system.

\bibliography{fusion_bib}

\end{document}